\documentclass[doublespacing]{elsart}

\usepackage{natbib, graphicx}

\usepackage{amssymb}
\journal{Physica A}
\begin{document}

\begin{frontmatter}



\title{Endo- vs. Exo-genous shocks and  relaxation rates
 in book and music "sales"}


\author[lambi]{R. Lambiotte},
\corauth[cor]{Corresponding author.} 
\ead{Renaud.Lambiotte@ulg.ac.be}
\author[lambi]{M. Ausloos},
\ead{marcel.ausloos@ulg.ac.be}

\address[lambi]{SUPRATECS, Universit\'e de Li\`ege, B5 Sart-Tilman, B-4000 Li\`ege, Belgium}

\begin{abstract}
In this paper, we analyze the response of music and book sales to an external field
and to buyer herding. We distinguish  endogenous and exogenous shocks. 
 We focus on some case studies, whose data have been collected from ranking on amazon.com. 
 We show that an ensemble of equivalent systems quantitatively respond in a similar way 
 to  a similar ''external shock'', indicating roads to universality features. In contrast to
 Sornette et al. [Phys. Rev. Lett. {\bf 93}, 228701 (2004)] 
 who seemed to find power law behaviors, in particular at long times, - 
 a law interpreted in terms
 of an epidemic activity, we observe that
  the relaxation process can be as well seen as an exponential one that saturates
  toward an asymptotic state, itself different from the pre-shock state. 
  By studying an ensemble of 111 shocks, on books or records, 
  we show that exogenous and endogenous shocks
   are discriminated by their $short-time$ behaviour: the relaxation time 
   seems to be twice shorter in endogenous shocks than in exogenous ones.
   We interpret the finding 
   through a simple
   thermodynamic model with a dissipative force.
   
\end{abstract}


\end{frontmatter}

\section{Introduction}

The  fluctuation-dissipation theorem is a corner stone of  Statistical Mechanics 
\cite{nyquist,kubo}.
Indeed, it allows to relate quantitatively two classes of 
dynamical features of macroscopic systems, namely fluctuation 
phenomena which are stochastic deviations from some equilibrium state, 
and the dissipative response of the system to an external field. 
 In out-of-equilibrium situations, however,
  such relations are not usually applicable {\em a priori}, even though they
   may be generalized in some particular cases \cite{barrat,dufty}.  
   A revival of studies has been recently seen on this theorem, 
   associated to features of shocks in sociological and economical networks \cite{sornette}.  Indeed, these systems are not usually considered to be mechanistic in essence, whence driven by deterministic Hamiltonian-like equations. They are not at equilibrium in the classical sense, and are usually subjects to outliers, like bubbles and crashes, in financial markets \cite{marcel}, the emergence of trends in sales and weblogs \cite{glance}, or epidemia and avalanches in opinion formation \cite{holyst}, music \cite{lam1,lam2} and science. Generally, these critical events may be caused by two kinds of mechanisms. On the one hand, there is the  response to some external field. In this case, one speaks of an exogenous shock. On the other hand, there is the spontaneous evolution of the system through a hierarchy of avalanches of all sizes. These extreme events are considered to be  endogenous, as it has been formalized by the theory of Self-Organized Criticality (SOC) \cite{bak}. In most non-physical systems, it is not easy to distinguish these two kinds of features since systems are usually  driven by an interplay of the two mechanisms.

 In the instance hereby considered, i.e. the case of marketing, the 
 sales of a product are usually driven by a reputation cascade as well
  as advertisements for the product. For example, in book sales dynamics, 
  some books reach their peak abruptly, followed by  decreasing sales, while 
  others reach their top rankings after a longer time on the market,  followed 
  by gradually falling sales. Sornette et al. \cite{sornette2} introduced an 
  epidemic-like model containing a long memory process for the buyers, 
  characterized by an exponent  $\theta \in[0,\frac{1}{2}]$. Endogenous shocks
   were shown to be formed by a slow increase of the sales, followed by a symmetric
    relaxation, i.e. the formation and the relaxation process behave 
    like $|t_c+t|^{2 \theta - 1}$, where $t=0$ corresponds 
    to the peak maximum and $t_c$ is an additional (not interpreted) positive parameter. 
    In contrast, an exogenous shock occurs abruptly and the sales decay faster, 
    like  $(t_c+t)^{\theta - 1}$. This prediction agrees with the intuition that 
    an endogenous shock finds its origin in the structure of the buyers network
     and should have a longer life time than a shock which has been imposed by
      an external cause to the system. By measuring relaxations of a large number of 
      endogenous and exogenous
      shocks, Sornette et al. found an average value of the  exponent $\theta\sim0.3 \pm 0.1$.
      However, the
       values of $t_c$, i.e. the short-time behaviour of the relaxation, are not considered.

 Nevertheless, despite these pioneering results, many fundamental questions 
 remain open in order to fulfill the original purpose of these studies, 
 namely a generalization of linear-response theory to marketing and sociological systems.
  In this respect let us point to Groot \cite{groot}   studies of
   sales data in a commodity market 
  from the point of view of the fluctuation spectrum and noise correlations.  
 
 After some caveat on
 the methodology inherent to such a type of studies and a visual discrimination between an ''endogenous'' and an ''exogenous'' shock on a case study, in Sect. 2,
 the present paper focuses on two important issues.
 On the one hand, we verify a required condition for the applicability of a 
 macroscopic description, namely we check the $reproducibility$ of the ''experiment'',
 both in music sales and on book sales in Sect. 3. This is done through two case studies.
 Thereby, we
 study the relaxation of equivalent systems to an external shock. 
  On the other hand, we revisit
  the study of Sornette et al. and focus on the $short$ time scale after a 
  sales maximum ($\sim$ 1 month). On this time scale, most of the systems rather
  show an exponential relaxation
   that was hidden by the parameter $t_c$, and not
   a power law decay. Moreover, we question whether the observed long time power-law 
   relaxations are not in fact associated to a saturation effect, in Sect. 4.
   Finally in Sect. 5, we  study the relaxations 
   of a ''large ensemble'' of shocks, characterized by their relaxation time $t_R$,
   and highlight a quantitative difference
    between exogenous 
   and endogenous shocks, whence in the short time range. The values of the 
   relaxation rates allow to discriminate rapidly between endo- and exo-genous shocks. 
   A theoretical model  based on simple thermodynamics
   taking into account a dissipative force has two easily measurable parameters given
   by the initial and asymptotic ranking states. The relaxation time 
   has a much more precise meaning than $t_c$.

 \section{Methodology}
 
 Amazon (www.amazon.cm) 
 is  the largest online store selling many goods, such as electronic devices, books, 
  or music albums. Among its descriptions of the product, the website assigns a rank
   which takes into account the number of copies that have been sold in the past. 
   The reverse translation of this rank into the number of sold copies  is not an easy task. 
   However, as discussed in ref., one may approximate this relation  by the  power 
   law $S \sim R^{-1/2}$, where $S$ is the number of sold items, 
   and $R$ is its rank in the Amazon database.
  Some warning is needed at once. The method for providing a rank to an item is 
  officially the following. For the top 10 000 best sellers from amazon.com sales., 
  the rank is updated each hour and takes into account the sales of the preceding 24 hours. 
  The next 90 000 ranks are updated daily, while the rest of the items is updated monthly
   with several different rules. In order to avoid such changes in the rank assignment
    method, and therefore artificial consequences on the time evolution of $R$, we 
    restrict our analysis to items that remain in the  $[1:100 000]$ interval.
 
\begin{figure}

\includegraphics[angle=-90,width=5.00in]{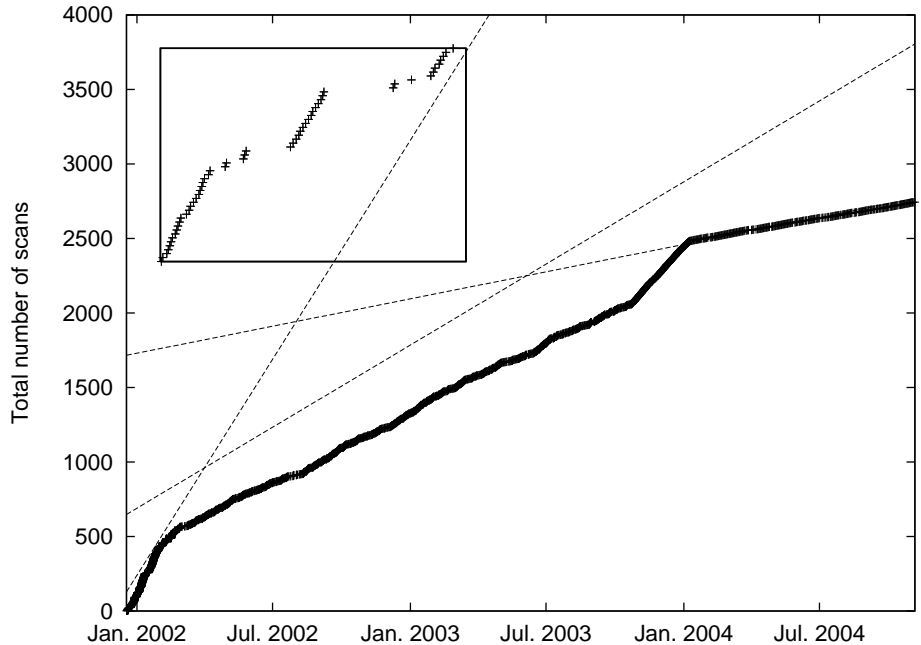}

\caption{\label{smithtime} Total number of scans as 
a function of time for the music album {\em XO} of Elliott Smith. The 
dash line corresponds to a fixed rate of 1 scan, 3 scans and 8 scans per day. 
In the inset, we plot a zoom of this curve in an  interval of 9 days in February 2002.}
\end{figure}

 In order to get the time evolution of $R$ over a long time range, 
 we have used data collected by junglescan (www.junglescan.com). 
 This website allows the users to enter a product's URL, and scans 
 its rank $R$ from the amazon website in the course of time. The time evolution 
 of the rank is then stored and accessible. One should note, however, that the
  scanning rate is not a constant. This is illustrated in figure \ref{smithtime}, 
  for a typical case where we plot the total number of scans for some music album, 
  the $XO$ of Elliott Smith,
   as a function of time. It is shown not only that the average scanning rate has
    evolved in the course of time, 1/day, 3/day, 8/day, but also
     that drastic local time interval changes may occur, as
     illustrated in the inset. Moreover, {\it due to unrelated to our study technical problems 
     in the junglescan server}, most of the scans that we have elaborated have been stopped
      at the beginning of November 2004.
 
\begin{figure}

\includegraphics[angle=-90,width=5.00in]{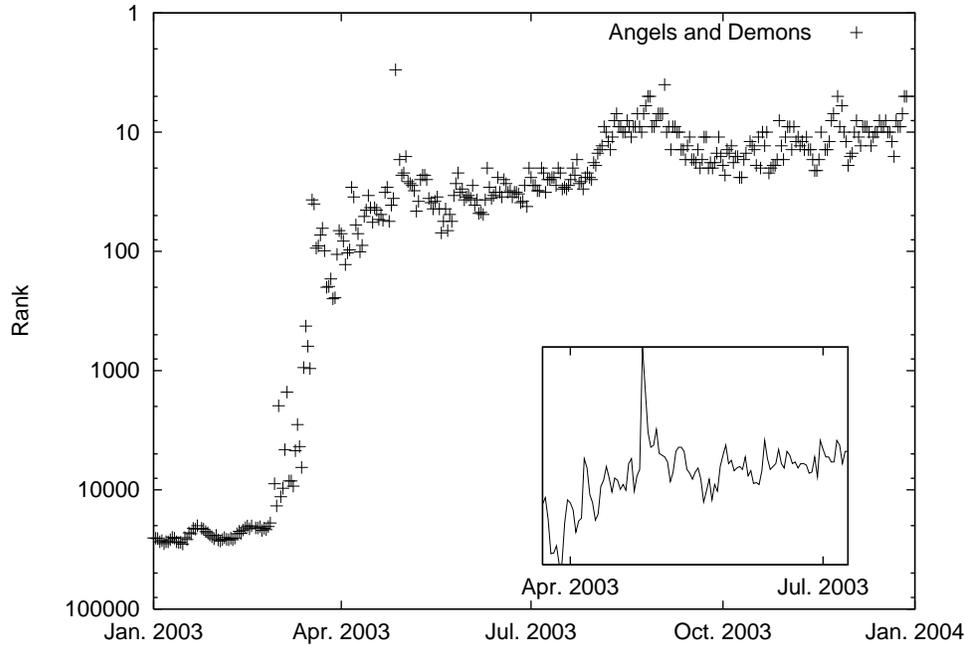}

 \caption{\label{browntime} Time evolution of the amazon.com rank of  {\em Angels and Demons}, written by Dan Brown. 
 In the inset, we zoom on the ranking during April 2003. Around September 2003, 
 there has been a slowly evolving bubble in book sales, - associated to an endogenous shock}
\end{figure}

 Let us now focus on a case study, namely {\em Angels and Demons} by Dan Brown.
  It has been first published on July  2001 and has been scanned from Oct. 1, 2002 
  till Nov. 4, 2004.  In  figure \ref{browntime}, one observes that, before March 2003, 
  {\em Angels and Demons} was not under the spotlight and that its rank remained 
  near the 30 000 rank. In March 2003, there is a qualitative change obviously
   associated with the publication and fame of {\em The Da Vinci Code}, which 
   makes the former book jumps in the top 100 after a few days.  This example shows
    the strong correlations between two book sales, namely {\em The Da Vinci Code}
     and {\em Angels and Demons}, abrupt changes in the book sales associated to what
      can be considered
     an
      $exogenous$ shock, around April 10th 2003, and an $endogenous$ shock
       in September 2003.
 
 \section{Experiment reproducibility}
 
 \begin{figure}

\includegraphics[angle=-90,width=5.00in]{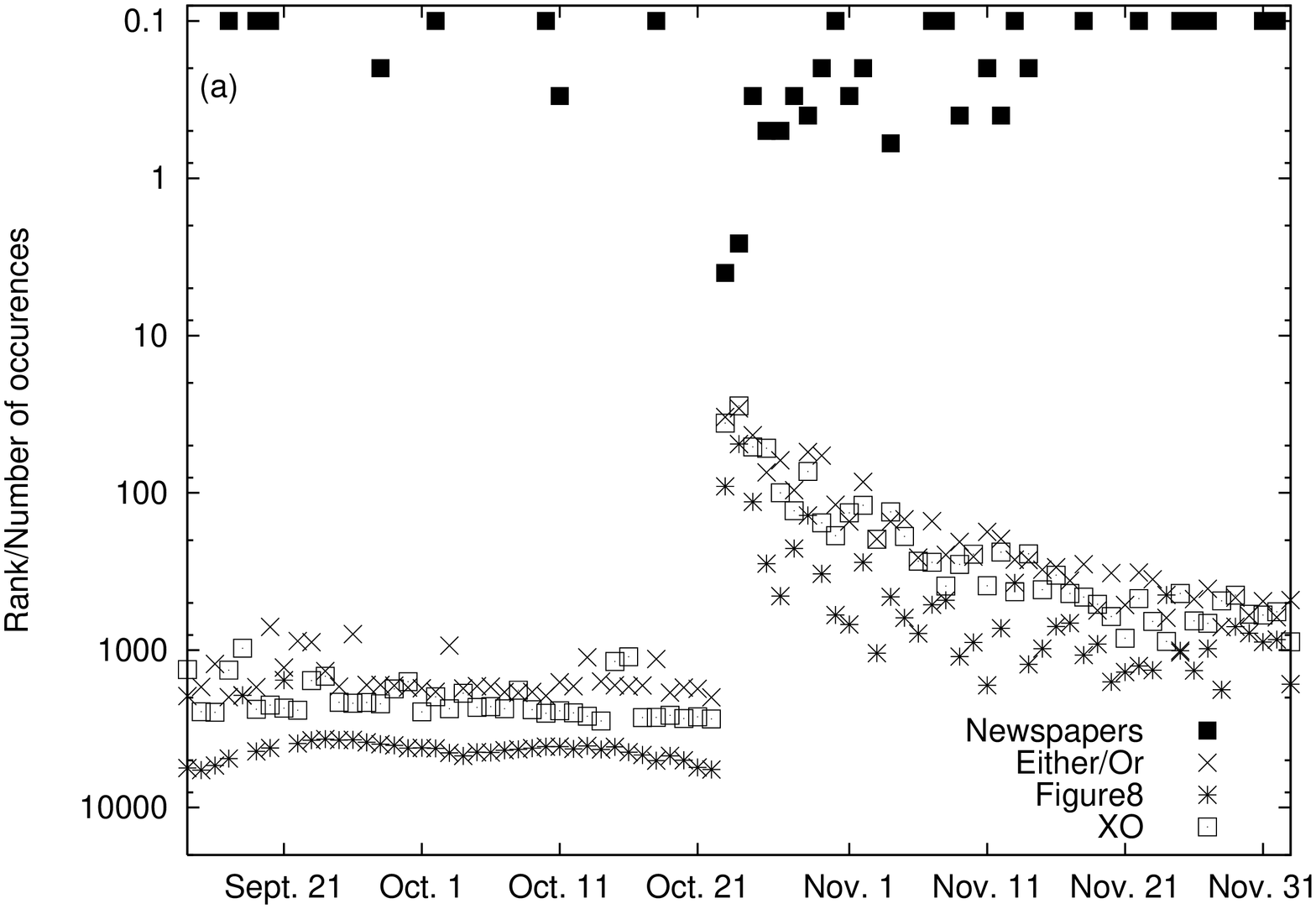}

\includegraphics[angle=-90,width=5.00in]{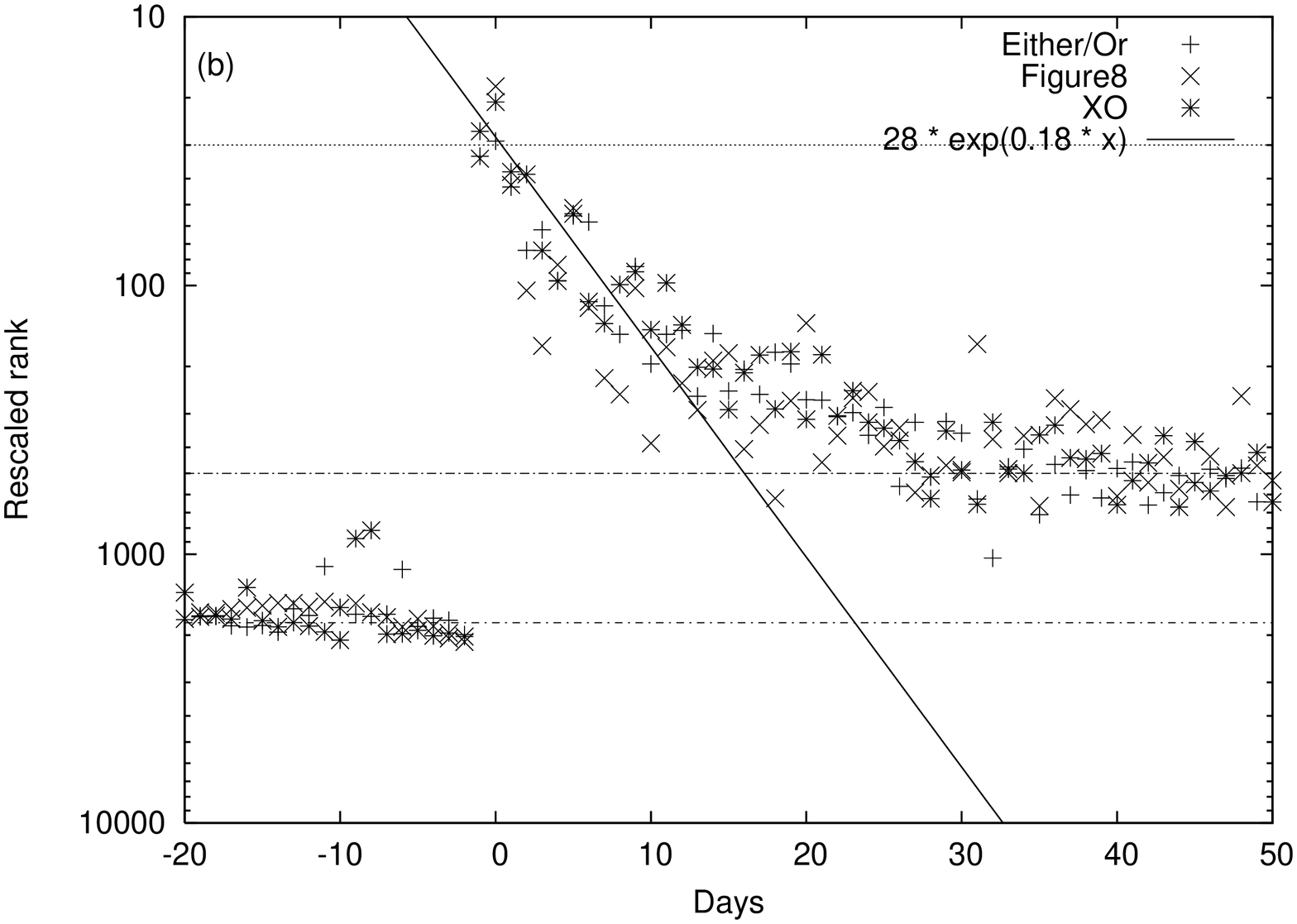}

\caption{\label{elliott} (a) Time evolution of the rank of 3 albums of Elliot Smith. 
The dark squares correspond  to the number of newspapers citing the artist,
 extracted from {\em www.highbeam.com}; (b) evolution 
 of the rescaled ranks of the albums in (a) around the exogenous shock. The dash 
 horizontal lines are guides 
 for the eye for the pre-shock, the shock, and the post-shock ranks }
\end{figure} 
  
 A first requirement in order to 
 apply a fluctuation-dissipation theorem to sales 
 is the existence of a well-defined macroscopic friction process in the system. 
 It implies that an ensemble of equivalent systems should evolve according 
 to that macroscopic law when they are put outside the stationary state. Of course,
  it is not easy to produce such a controlled experiment in the case of sales. 
  Nevertheless, we should verify this preliminary property through a detailed analysis
   of statistically appropriate cases.
 
  \subsection{Music sales}

  In order to do so, we consider  the sales of 3 albums of Elliott Smith: 
 {\em Figure 8}Ê(2000),
{\em XO}Ê(1998) and
{\em Either/Or}Ê(1998) (Fig. \ref{elliott}). On Oct. 22, 2003, this young folk music writer
 died from an apparent suicide. The next day, in response to this unexpected focus
  around his personality, all of his albums underwent an abrupt jump of their sales 
  that relaxed over a few weeks. This is a perfect example of exogenous shock, whose 
  accidental source moreover allows to get rid of any marketing strategy. One should 
  also note, as illustrated in figure \ref{elliott}, that the event produced a very 
  localized stream in the media, on the day of his death and the next one. This allows 
  to affirm that the slow relaxation process of the sales obviously finds its origin 
  in the sales dynamics itself. 
  
 Let us stress that this (external) shock makes the system 
  reach a stationary state different from the pre-shock state. In the case of
   {\em Either/Or}, for instance, the pre-shock state fluctuates around rank $1800$. 
   The shock makes the rank jump to $17$, followed by a relaxation of some weeks to
    an asymptotic state around rank $500$. This behaviour is similar to most other 
      cases that we have studied, namely an exponential relaxation for short times 
      that saturates 
    to an asymptotic value, i.e. the post-shock state. We discuss this issue further 
    in the remainder of this section.

The  three album signals  obviously differ from each other (Fig. \ref{elliott}). 
This is well seen in the pre-shock regime where each album average rank
 ranges between 1 800 and 8 000. A rescaling would account for emphasizing 
 the qualitative differences between the albums. Assuming a constant in
  time scaling factor, the average pre-shock values become almost equal, 
   in so doing indicating universality features. Therefore, we observe that the three
   data collapse on each other, and that the short time rank relaxation behaves 
   like $e^{\lambda t}$, with a relaxation coefficient
    $\lambda=0.18$. This means that the three equivalent systems respond 
    in the same way to the studied external shock.

\subsection{Book sales}

\begin{figure}

\includegraphics[angle=-90,width=5.00in]{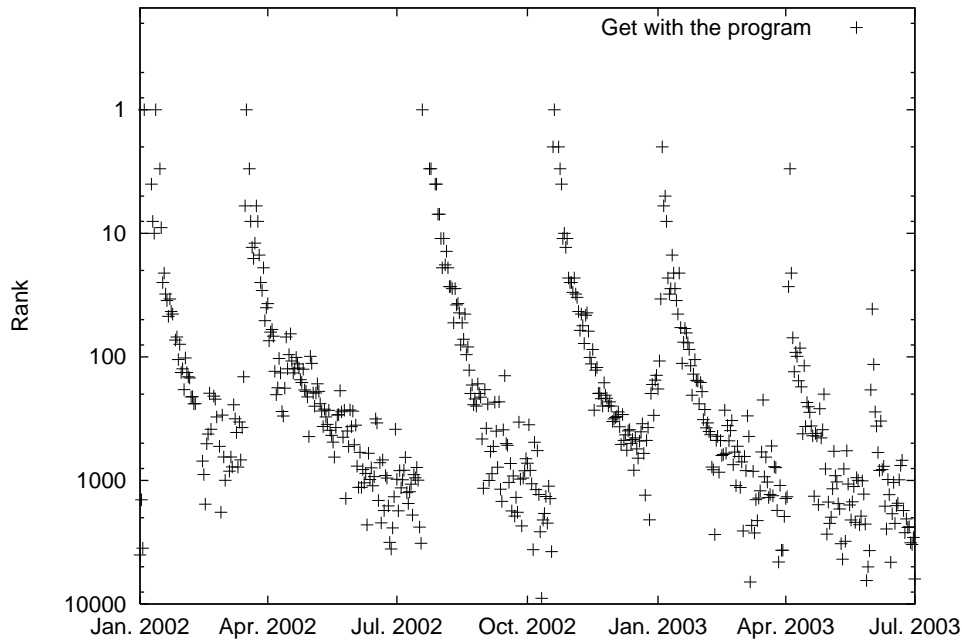}

\caption{\label{oprah} Time evolution of the ranks of {\em Get with the program}
 by B. Greene  }
\end{figure}

\begin{figure}

\includegraphics[angle=-90,width=5.00in]{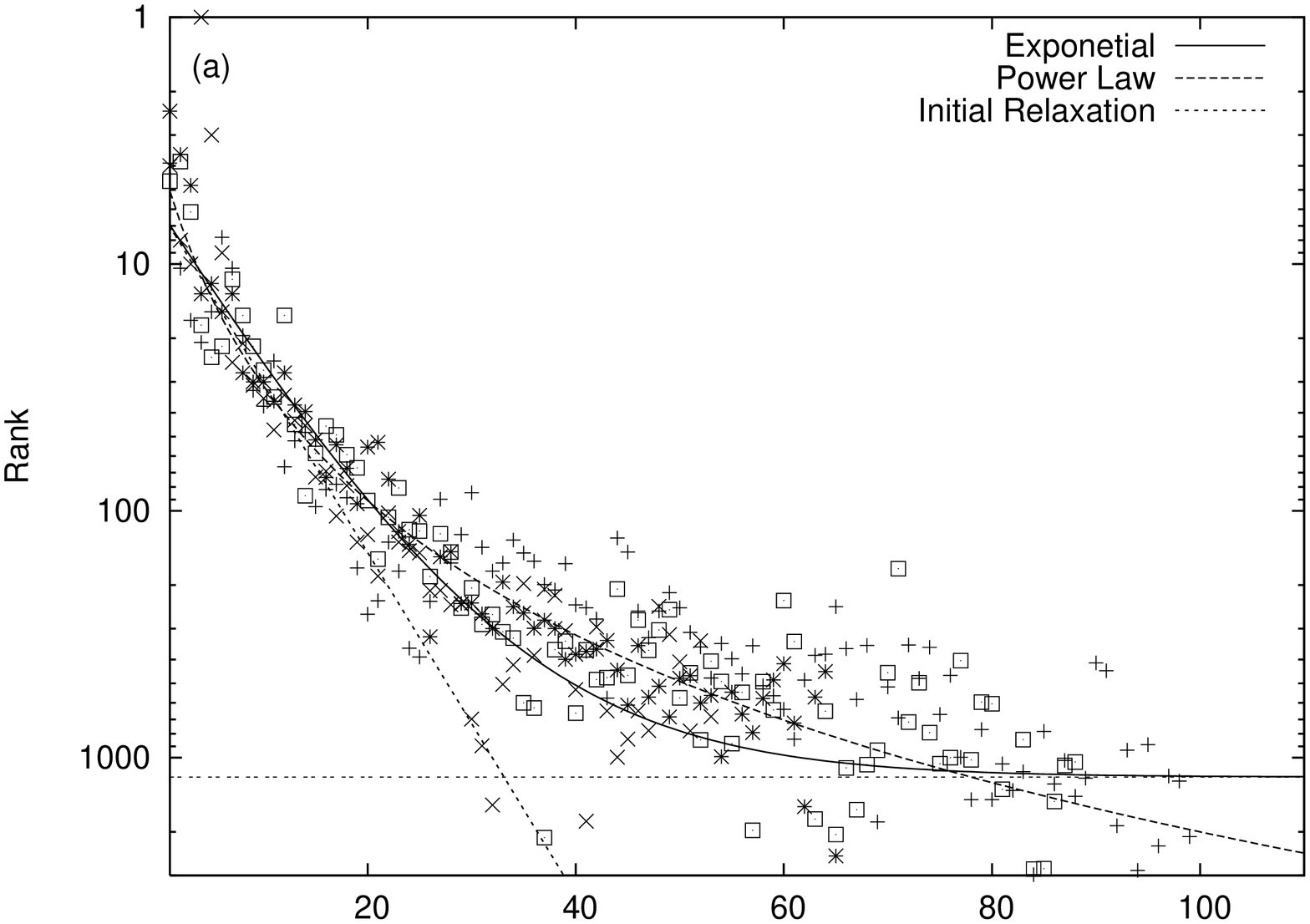}

\includegraphics[angle=-90,width=5.00in]{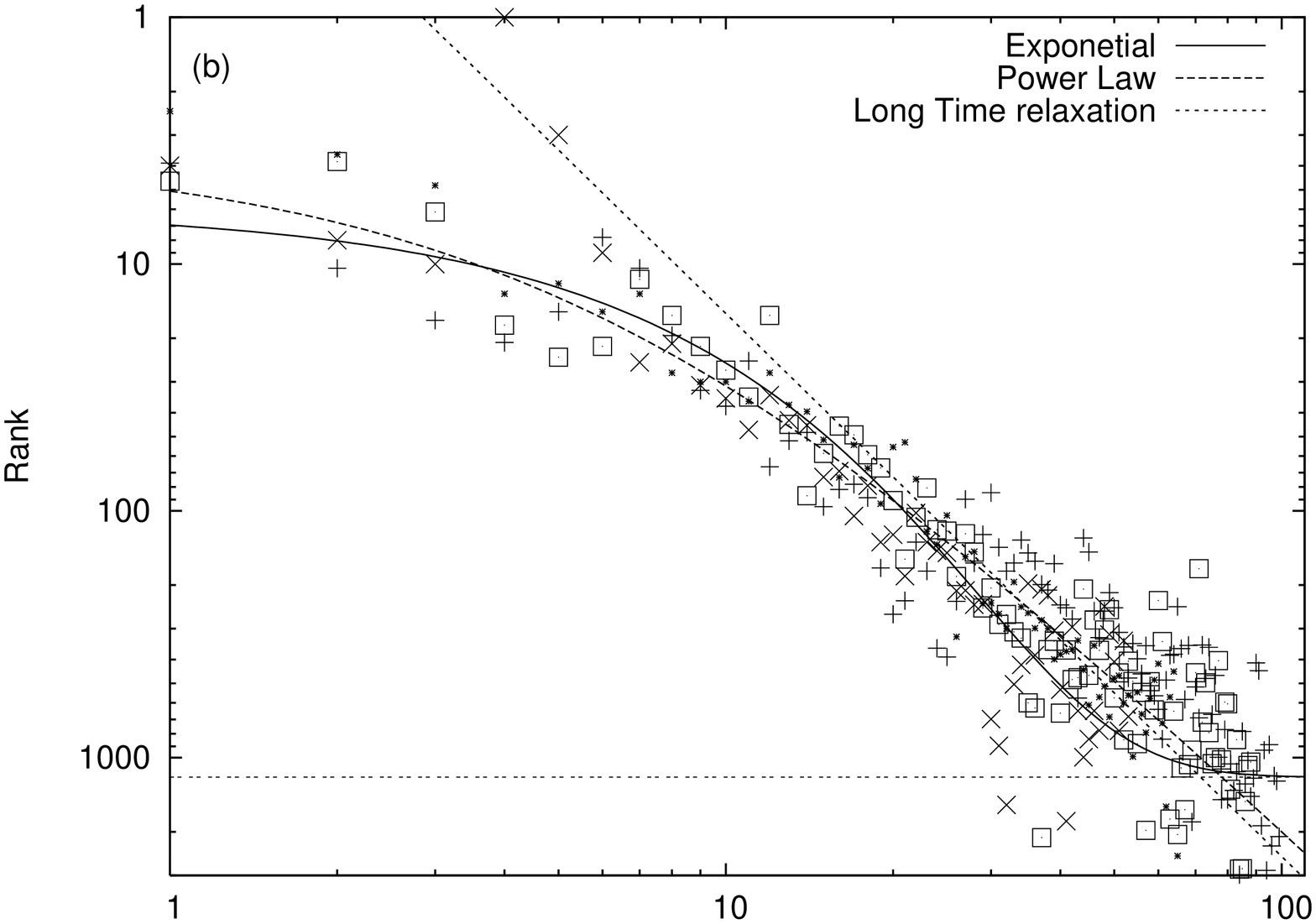}

\caption{\label{greene}We focus on four exogenous shocks of {\em Get with the program},
 in (a) log-normal and (b) log-log scale. The solid lines represent the exponential 
 relaxation  Eq.(\ref{formula}), with $R_{\infty}=1200$, $R_{0}=6$ and $\lambda=0.16$,
  and the power law $0.07 ~ (6+t)^{2.2}$. We also plot the short time and long time
   asymptotic behaviours of these functions, namely $R= 6 ~ e^{0.16 t}$ and $R \sim t^{2.2}$}
\end{figure}

Another way to verify the reproducibility of an experiment is 
to consider the response of one system to several "equivalent" shocks.
 To do so, we have considered the rank of  {\em Get with program}, 
 written by B. Greene, and focused of the jumps due to his frequent 
 passages at the Oprah Winfrey Show, between Jan. 02 and Jul. 03 
 (see Fig. \ref{oprah}). The rank relaxations of the first 4 shocks are plotted 
 in figure \ref{greene}, under (a) log-normal  and (b) log-log scale. 
 Data analysis leads to the same conclusion as in the previous (music sales) example, namely an initial exponential relaxation followed by saturation,
  and confirms that book and music sales are similar macroscopic phenomena, 
  with respect to shocks.

\section{Short time and long time behaviours}

\begin{figure}

\includegraphics[angle=-90,width=5.00in]{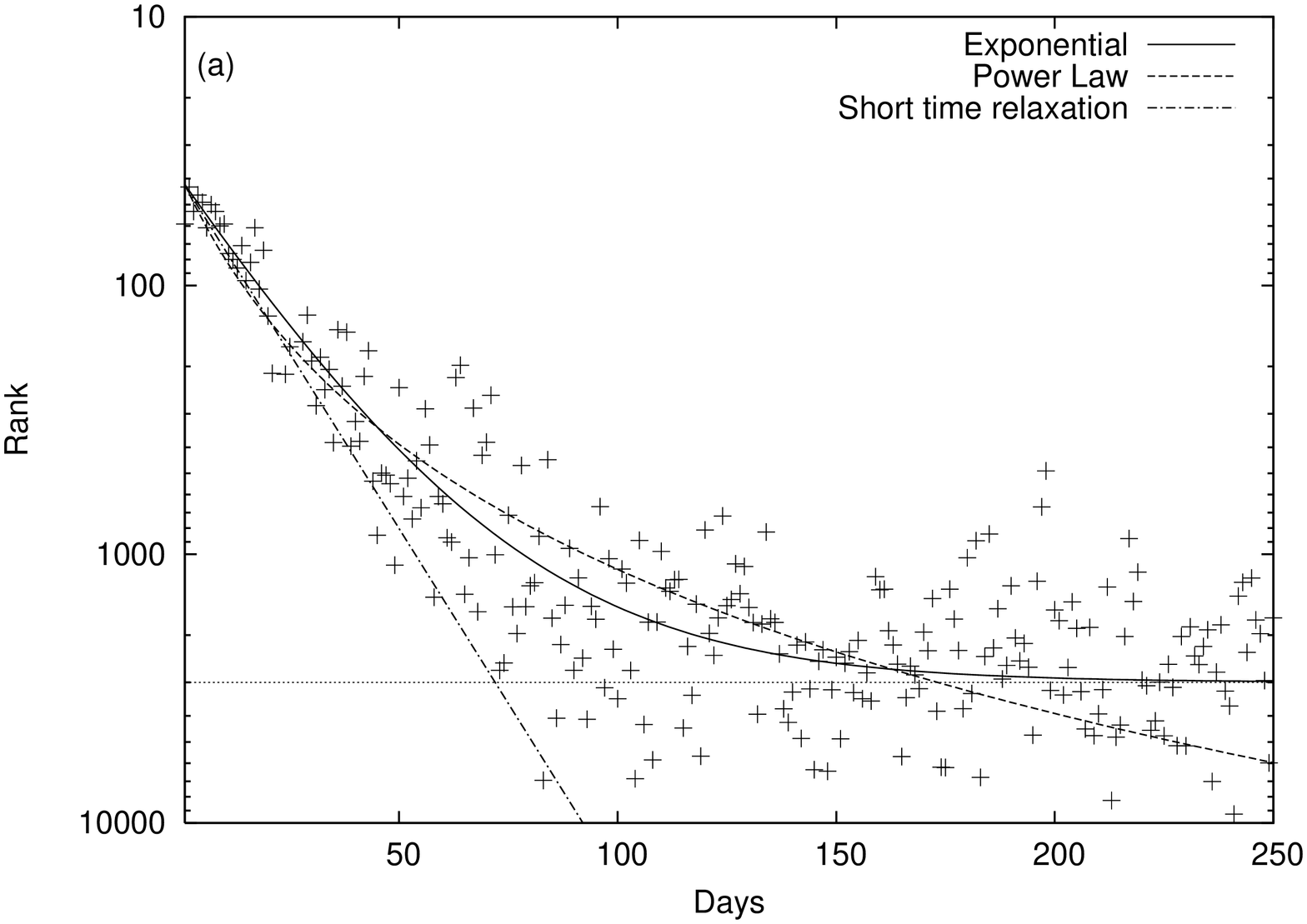}

\includegraphics[angle=-90,width=5.00in]{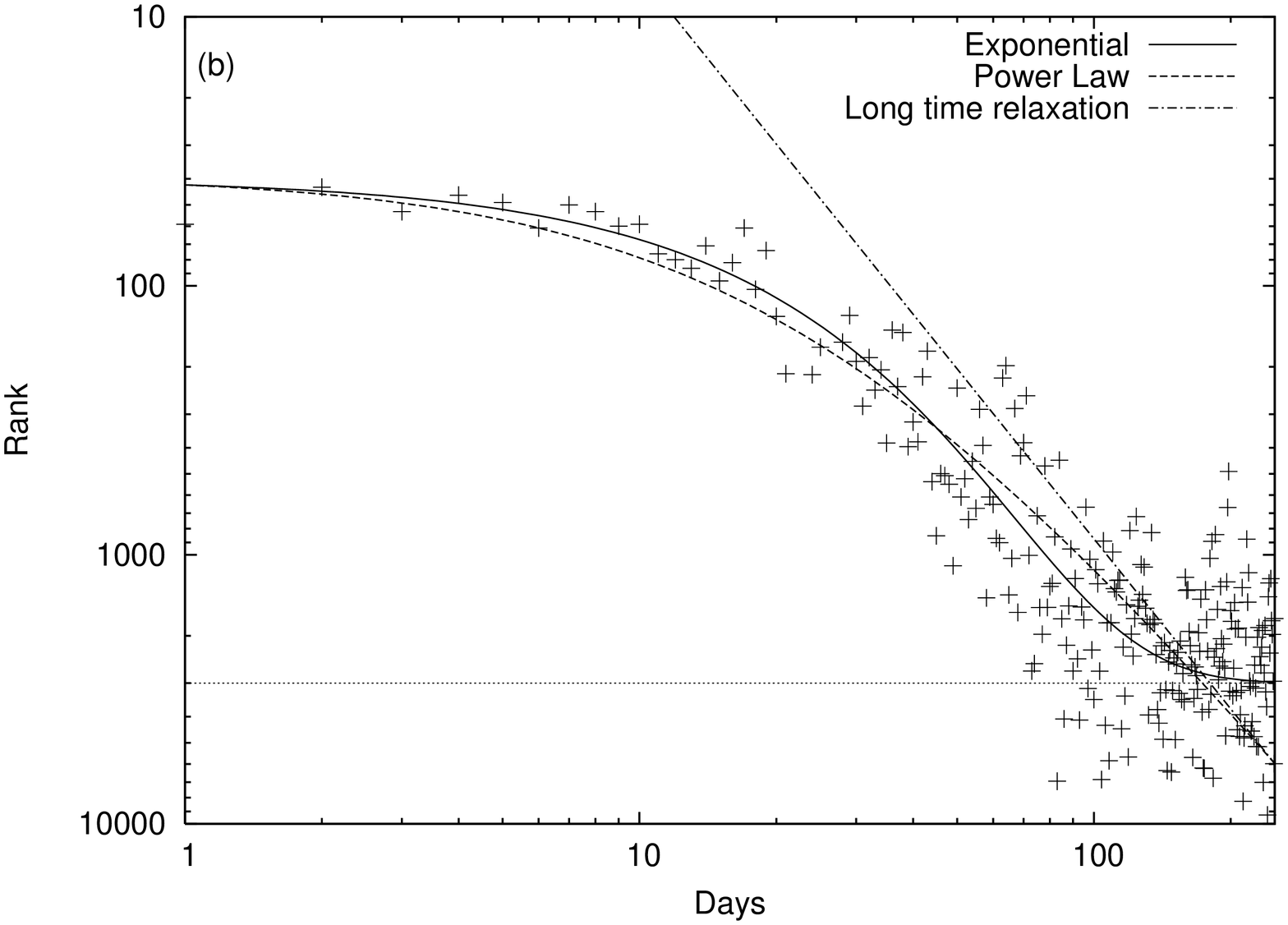}

\caption{\label{long} We focus on the endogeneous peak of
 {\em Heaven and Earth} by N. Roberts, in (a) log-normal and (b) log-log scale.
  The solid lines represent the exponential relaxation \ref{formula}, 
  with $R_{\infty}=3000$, $R_{0}=40$ and $\lambda=0.06$, and the power
   law $0.045 ~ (25+t)^{2.1}$. We also plot the short time and long time
    asymptotic behaviours of these functions for comparison, namely $R= 6 ~ e^{0.06 t}$ and $R \sim t^{2.1}$. }
\end{figure}

The above results   suggests to question the universality of
the power law, and to emphasize the short time scale
behavior relaxation, i.e. to draw a parallel
   between the relaxation coefficient $\lambda$ and a dissipative force.
Moreover since
 Sornette et al. \cite{sornette2} found relaxation processes of book sales 
 governed by  power-laws $(t_c+t)^{\mu}$, it is of interest to compare
  this power-law behaviour to the exponential one found in the previous section. 
 Notice that these authors focused their analysis on the long time behaviour. However, 
there was no detailed analysis of the short-time scale of the relaxation in their work, 
nor a clear explanation of the   parameter $t_c$. The case studies in the previous
 section suggest that the short-time relaxation processes  are rather exponential.
In the following, we present a simple
alternative description of the relaxation that accounts
 for the short-time dissipation, as well as for the asymptotic saturation 
 leading to the asymptotic state.

First, let us assume that the sales can be related to the rank through the relation
 $S=R^{- \gamma}$, $\gamma=\frac{1}{2}$. Let us stress that the exact value of
  this exponent is not critical, and that this value was chosen only in order to allow
   comparisons with the results of Sornette et al. Then, we assume that the response 
   of the sales to a shock is an exponential decrease toward an asymptotic state
    different of zero. 
The non-vanishing asymptotic state is, in the thermodynamic sense, due to the
 continuous agitation of the buyers, agitation that encompasses internal dynamics 
 due to buyers interactions and small external kicks. The simplest form is

\begin{equation}
S= S_{\infty} + (S_0-S_{\infty}) e^{-\frac{\lambda}{2}t}
\end{equation}
that leads to the following expression for  the item rank.
\begin{equation}
\label{formula}
R= (R_{\infty}^{-\frac{1}{2}} + (R_{0}^{-\frac{1}{2}}- R_{\infty}^{-\frac{1}{2}}) 
e^{-\frac{\lambda}{2}t})^{-2}
\end{equation}
This expression reduces to $R \sim R_{0} ~ e^{\lambda t}$ in the small time limit.
Therefore, this description has the advantage to depend directly on observable quantities,
 namely the friction coefficient $\lambda$, and both initial and asymptotic values of the 
 ranks. 

In figures \ref{greene} and \ref{long}, we apply this fitting procedure to an exogenous 
({\em Get with the program})
and an endogenous ({\em Heaven and Earth}, by N. Roberts) shock.
Moreover, we compare the results  with the
 power-law fits $(t_c+t)^{\mu}$ of Sornette et al. 
 In the first case (Fig.\ref{greene}), the fitted parameters of  (\ref{formula}) are $R_{\infty}=1200$, $R_{0}=6$ and $\lambda=0.16$,
  and the power law fit is  $0.07 ~ (6+t)^{2.2}$.
    In the other case (Fig. \ref{long}), the parameters of  (\ref{formula}) are  $R_{\infty}=3000$, $R_{0}=40$ and $\lambda=0.06$, and
  the  power
   law  $0.045 ~ (25+t)^{2.1}$. In these figures, we also plot the short time 
   and long time
    asymptotic behaviours of these functions for comparison. 
 One observes that both approaches lead to
  similar results that can not be discriminated given the data accuracy.  This is verified by focusing on the mean square deviation:
  \begin{equation}
   \sigma_F=\sqrt{\frac{\sum_{i=1}^K(\log(R^i) - \log(R_F^i))^2}{K}}
   \end{equation}
    where $K$ is the number of data points, $R^i$ the rank from the data set and  $R^i_F$ the value of the fitted function, either exponential ($F=E$) or power law ($F=P$).
  In the case of {\em Heaven and Earth}, for instance, these values for the exponential and the power law fit 
  are very close,  $\sigma_E=0.249$ and $\sigma_P=0.278$.
  This equivalence is remarkable, given the long time intervals considered 
  hereby ($100$ and $200$ days).  However, one should   note the very high 
  value of $t_c=25$ in figure \ref{long}, that dominates the power-law decrease 
  over a long time scale, and thereby might mask  the extension of the
  exponential relaxation at long times in Sornette et al. approach. It is also fair to recognize the large but similar values of the 
  exponent $\mu$.

A comparison of the friction coefficient $\lambda$ in Figs. \ref{greene} and \ref{long}
 suggests that exogenous and endogenous shocks occur on different time scales
  $t_R \equiv \lambda^{-1}$. One should stress here that this assertion is different 
  from that Sornette et al., i.e. a discrimination of shocks can be  based on the
   short-time behaviour of the relaxation process, and not on its tail.

\section{Universality behaviours}

\begin{figure}

\includegraphics[angle=-90,width=5.00in]{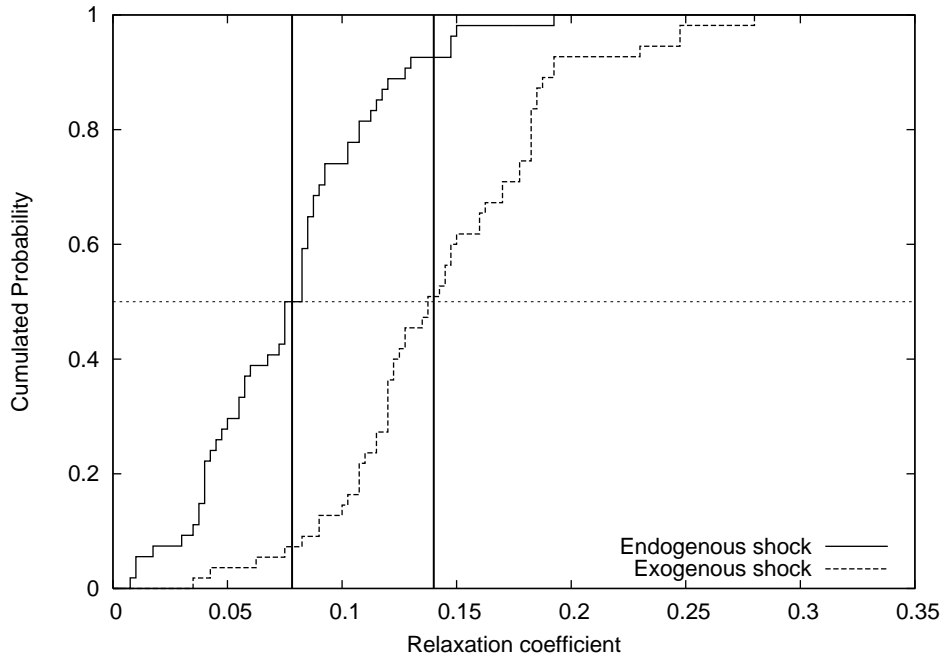}

\caption{\label{histo} Cumulated probability of the relaxation coefficient
 $\lambda$, in the case of exogenous and endogenous shocks.  The vertical
  lines point to the average values  $0.07$ (endogenous) and $0.14$ (exogenous), 
  that correspond to the characteristic relaxation times $t_R \sim 13
$ and $t_R \sim 7$ days respectively}
\end{figure} 

In order to highlight this observation, we have focused on 111 
(56  endo- and 55 exo-genous) shocks 
extracted from the junglescan data, indiscriminating between books and records. 
The shocks were found by coarse-graining the signal over one week, and by neglecting
 relaxations with $\lambda<0.01$, such that shocks are distinguished from quick and large 
 fluctuations. Moreover, we have only considered shocks occurring at least 1 month after
  the begin of the scans, i.e. in order to reject   new products and dumping acts,
   and whose minimum
   rank verified $R_0<100$.  Finally, exogenous and endogenous shocks were 
   visually discriminated by
    focusing on the pre-shock acceleration.
    
In order to measure $\lambda$, we have looked for the best fitting exponential
 in an interval of 15 days after the shock. The resulting cumulated histograms for $\lambda$ 
 are plotted in  figure \ref{histo}. The histograms for the exogenous and the endogenous
  shocks are  markedly different. The peak and the average  
  probability distribution of $\lambda$ of the endogenous shocks are well separated
  from those of the exogeneous ones :
   $<\lambda>_{exo}  \sim 2 <\lambda>_{endo}=0.14$. 
 This obviously confirms that the initial decay of exogenous and endogenous 
shocks occurs on different time ranges. 

 \section{Conclusion}

We have examined endogenous and exogenous so called shocks in music and book sales, 
measured from their rank in amazon.com.
 We have focused on some case studies. 
 We have shown that music and book sales  quantitatively respond in a similar way 
 to  a similar ''external shock''. In contrast to
 Sornette et al.  \cite{sornette2}
 who   found power law behaviors, and
  interpreted the finding in terms
 of an epidemic activity, we have observed that
  the relaxation can be   seen as an exponential that saturates
  toward an asymptotic state, itself different from the pre-shock state. 
  We have emphasized the non universal value of $t_c$ and found power law
  exponents quite different  from Sornette et al., both larger but very similar for
  the two types of shocks.
  We prefer to interpret our findings
   through a simple macroscopic 
     model with agitated herding buyers and  a dissipative force.
  By studying an ensemble of 111 shocks, on books or records, 
  we have shown that exogenous and endogenous shocks
   are discriminated by their $short-time$ behaviour: the relaxation time $t_R$
   seems to be twice shorter in exogenous shocks than in endogenous ones.
   This is a relevant (scientific and economic) result that completes 
   the discrimination procedure of 
Sornette et al.  and should be verified in other fields related to 
trend emergence, such as opinion formation, financial bubbles or scientific avalanches, 
on various networks
indicating roads to universality classes.

{\bf Acknowledgements}

Part of this work results from RL financing through the ARC
(02-07/293 )
and the CREEN  (012864) project, which MA also thoroughly acknowledges.


\begin{thebibliography}{0}

\bibitem{nyquist}  
H. Nyquist, {\em Phys. Rev.}, {\bf 32} (1928) {110} 

\bibitem{kubo}  
R. Kubo, {\em J. Phys. Soc. Japan}, {\bf 12} (1957) {570} 

\bibitem{barrat}
G. D'Anna, P. Mayor, A. Barrat, V. Loreto, F. Nori, {\em 
Nature}, {\bf 424} (2003) 909

\bibitem{dufty}
J. W. Dufty, V. Garzo, {\em J. Stat. Phys.}, {\bf 105} (2001) 723

\bibitem{sornette}  
D. Sornette, Monograph on extreme events,  Jentsch editor (Springer, 2005) 

\bibitem{marcel}
M. Ausloos, P. Clippe, A. Pekalski, {\em Physica A}, {\bf 332} (2004) 394

\bibitem{glance} 
L. Adamic, N. Glance,
www.blogpulse.com/papers/2005/AdamicGlanceBlogWWW .pdf

\bibitem{holyst}
J. Holyst, K. Kacperski, F. Schweitzer
{\em Physica A}, {\bf 285} (2000) 199

\bibitem{lam1}
R. Lambiotte and M. Ausloos, physics/0508233

\bibitem{lam2}
R. Lambiotte and M. Ausloos, physics/0509134

\bibitem{bak}  
P. Bak, C. Tang, K. Wiesenfeld {\em Phys. Rev. Lett.}, {\bf 59} (1987) {381} 

\bibitem{sornette2}  
D. Sornette, F. Deschatres, T. Gilbert, Y. Ageon {\em Phys. Rev. Lett.},
 {\bf 93} (2004) 228701 

\bibitem{groot}  
R.D. Groot, L\'evy distribution and long correlation times in supermarket sales,
eprint arXiv:cond-mat/0412163


\end{thebibliography}
\end{document}